\documentclass[fleqn,usenatbib]{mnras}
\usepackage{newtxtext,newtxmath}
\usepackage[T1]{fontenc}
\DeclareRobustCommand{\VAN}[3]{#2}
\let\VANthebibliography\thebibliography
\def\thebibliography{\DeclareRobustCommand{\VAN}[3]{##3}\VANthebibliography}
\usepackage{graphicx}
\usepackage{amsmath}	

\usepackage{amssymb}	

\newcommand{\ha}{\hbox{H$\alpha$}}
\newcommand{\hb}{\hbox{H$\beta$}}

\newcommand{\oiii}{\hbox{[O\,{\sc iii}]}}
\newcommand{\nii}{\hbox{[N\,{\sc ii}]}}
\newcommand{\sii}{\hbox{[S\,{\sc ii}]}}
\newcommand{\hii}{\hbox{H\,{\sc ii}}}

\title[MaNGA:properties of misaligned galaxies]{SDSS-IV MaNGA : spatial resolved properties of kinematically misaligned galaxies}
\author[Xu et al.]{\raggedright
Haitong Xu,$^{1,2,3}$
Yanmei Chen,$^{1,2,3}$\thanks{E-mail: chenym@nju.edu.cn}
Yong Shi,$^{1,2,3}$
Yuren Zhou,$^{1,2,3}$
Dmitry Bizyaev,$^{4,5}$
Min Bao,$^{1,2,3}$
\newauthor
Minje Beom,$^{6}$
Jos\'e G. Fern\'andez-Trincado$^{7,8}$ and
Xiao Cao$^{1,2,3}$
\\
$^{1}$School of Astronomy and Space Science, Nanjing University, Nanjing 210093, China\\
$^{2}$Key Laboratory of Modern Astronomy and Astrophysics (Nanjing University), Ministry of Education, Nanjing 210093, China\\
$^{3}$Collaborative Innovation Center of Modern Astronomy and Space Exploration, Nanjing 210093, China\\
$^{4}$Apache Point Observatory and New Mexico State University, PO Box 59, Sunspot, NM 88349-0059, USA\\
$^{5}$Sternberg Astronomical Institute, Moscow State University, Moscow 119992, Russia\\
$^{6}$Department of Astronomy, New Mexico State University, Las Cruces, NM 88001, USA\\
$^{7}$Instituto de Astronom\'ia, Universidad Cat\'olica del Norte, Av. Angamos 0610, Antofagasta, Chile\\
$^{8}$Universidad de Atacama, Copayapu 485, Copiap\'o, Chile
}
\date{Accepted XXX. Received YYY; in original form ZZZ}
\pubyear{2021}
\begin{document}
\label{firstpage}
\pagerange{\pageref{firstpage}--\pageref{lastpage}}
\maketitle

\begin{abstract}
    We select 456 galaxies with kinematically misaligned gas and stellar components from 9546 parent galaxies in MaNGA, and classify them into 72 star-forming galaxies, 142 green-valley galaxies and 242 quiescent galaxies. Comparing the spatial resolved properties of the misaligned galaxies with control samples closely match in the D$_n$4000 and stellar velocity dispersion, we find that: (1) the misaligned galaxies have lower values in $V_{\rm gas}/{\sigma}_{\rm gas}$ and $V_{\rm star}/{\sigma}_{\rm star}$ (the ratio between ordered to random motion of gas and stellar components) across the entire galaxies than their control samples; (2) the star-forming and green-valley misaligned galaxies have enhanced central concentrated star formation than their control galaxies. The difference in stellar population between quiescent misaligned galaxies and control samples is small; (3) gas-phase metallicity of the green valley and quiescent misaligned galaxies are lower than the control samples. For the star forming misaligned galaxies, the difference in metallicity between the misaligned galaxies and their control samples strongly depends on how we select the control samples. All these observational results suggest external gas accretion influences the evolution of star forming and green valley galaxies, not only in kinematics/morphologies, but also in stellar populations. However, the quiescent misaligned galaxies have survived from different formation mechanisms.    
\end{abstract}

\begin{keywords}
galaxies : evolution -- galaxies : kinematics and dynamics
\end{keywords}

\section{Introduction}

Galaxies can grow through both internal and external processes. The internal processes include stellar winds, supernova explosion, AGN feedback, secular evolution, and the external processes include mergers and gas accretion. Considering widely accpeted tidal torque theory (TTT), baryons acquire angular momentum from background gravitational tidal field when matter collapses to form a galaxy/halo \citep{1951pca..conf..195H, 1969ApJ...155..393P, 1970Ap......6..320D}, during this process angular momentum conserves, the newly formed stars inherit gaseous dynamical properties \citep{1978MNRAS.183..341W, 1980MNRAS.193..189F, 1998MNRAS.295..319M}. If a galaxy grows only through internal processes, the phenomena of galactic scale gas-star misalignment in velocity fields will not be as popular as we obseved (30$\sim$40\% in elliptical and lenticular galaxies). Thus galaxies with different distributions of velocity fields in gas and stellar components are the ideal laboratory to study the influence of external processes on galaxy evolution, whether they completely reshape the structure of the host galaxies or merely perturb them; whether the strength of influence depends on different physical properties or type of the host galaxies?

The phenomena of misaligned velocity fields between gas and stellar components have been observed ubiquitously in elliptical and lenticular galaxies with a fraction up to 30$\sim$40\% \citep{1992ApJ...401L..79B, 1996MNRAS.283..543K, kannappan_broad_2001, sarzi_sauron_2006, davis_atlas3d_2011, barrera-ballesteros_kinematic_2014, barrera-ballesteros_tracing_2015, chen_growth_2016, jin_sdss-iv_2016}, but much fewer (2$\sim$5\%) in star-forming galaxies \citep{chen_growth_2016, jin_sdss-iv_2016, bryant_sami_2019}. The lower fraction of the galaxies with decoupled gas and star kinematics in gas-rich star forming galaxies is due to that the collision cross-section between pre-existing and accreted gas is large enough to influence the retrograde angular momentum of the accreted gas. The decoupled gas-star kinematics only appears in galaxies where the angular momentum of the accreted gas is larger than the pre-existing gas. While for the gas-poor galaxies with low SFR, the accreted gas would survive for $\sim$1 to 5 Gyr \citep{davis_depletion_2016} since the interaction with existing gas is negligible.

In the last decade, especially since the development of large integral-field spectroscopic surveys, such as CALIFA \citep{2012A&A...538A...8S}, SAMI \citep{2012MNRAS.421..872C} and $\rm ATLAS^{3D}$ \citep{2011MNRAS.413..813C}, series of works have been developed in order to understand the origin of the kinematically misaligned gas components. These studies include morphologies/environments \citep{davis_atlas3d_2011, barrera-ballesteros_kinematic_2014, barrera-ballesteros_tracing_2015, jin_sdss-iv_2016, bassett_formation_2017, 2021MNRAS.505.5075Y}, stellar populations \citep{chen_growth_2016, jin_sdss-iv_2016, bevacqua_sdss-iv_2021}, gas-phase metallicities \citep{chen_growth_2016, jin_sdss-iv_2016}, kinematics \citep{katkov_decoupled_2014, naab_atlas3d_2014, bevacqua_sdss-iv_2021}. Following the development in the observations, gas-star misaligned galaxies are also found in cosmological simulation \citep{osman_strong_2017, bassett_formation_2017, taylor_origin_2018, duckworth_sdss-iv_2019, starkenburg_origin_2019, 2020MNRAS.492.1869D, 2020MNRAS.495.4542D, khoperskov_extreme_2020, khim_stargas_2021}, in which we can trace back the formation histories of the misaligned galaxies, searching for the physical mechanisms responsible for misaligned gas-star kinematics. Although several different mechanisms, such as gas precession and AGN feedback, have been proposed to produce or make it easier to produce misaligned phenomena, all of these works share a common sense that external processes, e.g. merging and gas accretion, play significant roles in the formation of misaligned gas-star components. In particular, \citet{lu_hot_2021} used the IllustrisTNG simulation \citep{marinacci_first_2018, naiman_first_2018, nelson_first_2018, nelson_first_2019, pillepich_first_2018, pillepich_first_2019, springel_first_2018} and found that the gas-star misaligned star-forming disc galaxies have experienced more frequent retrograde mergers (with respect to their stellar spins) throughout their history. During a retrograde merger, gas at the outskirts was first perturbed by the incoming galaxy, becoming misaligned with respect to the existing stellar spin. As gas was then accreted onto the central stellar disk, the galaxy would finally exhibit the gas-star kinematically misaligned feature.

\begin{figure*} 
    \includegraphics[width=2.0\columnwidth]{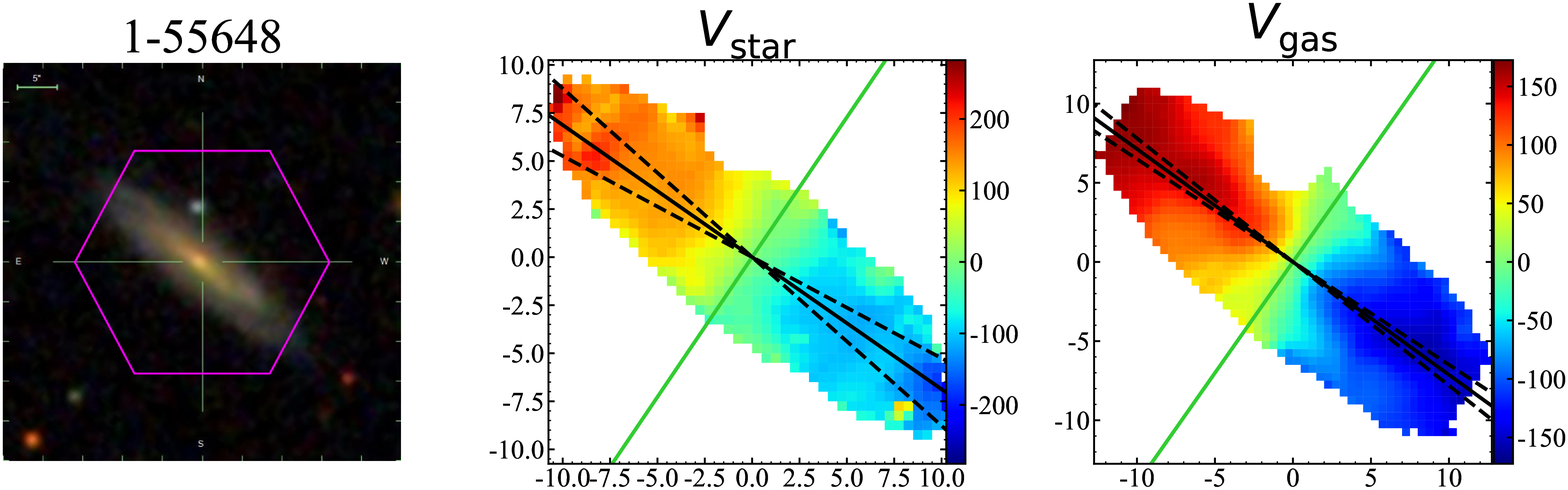}
    \includegraphics[width=2.0\columnwidth]{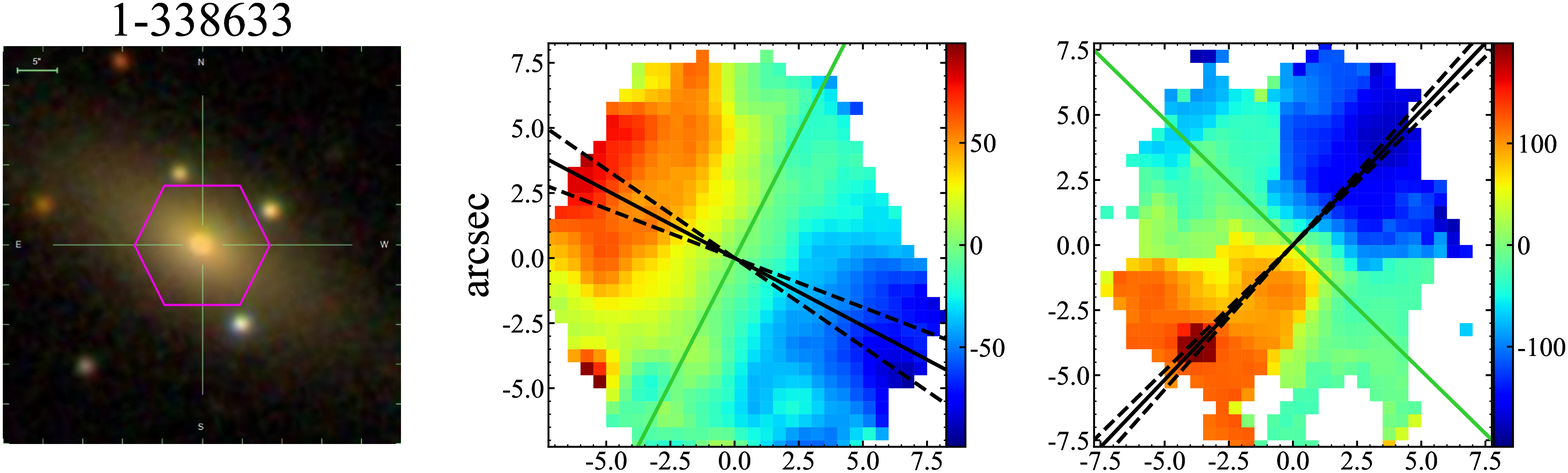}
    \includegraphics[width=2.0\columnwidth]{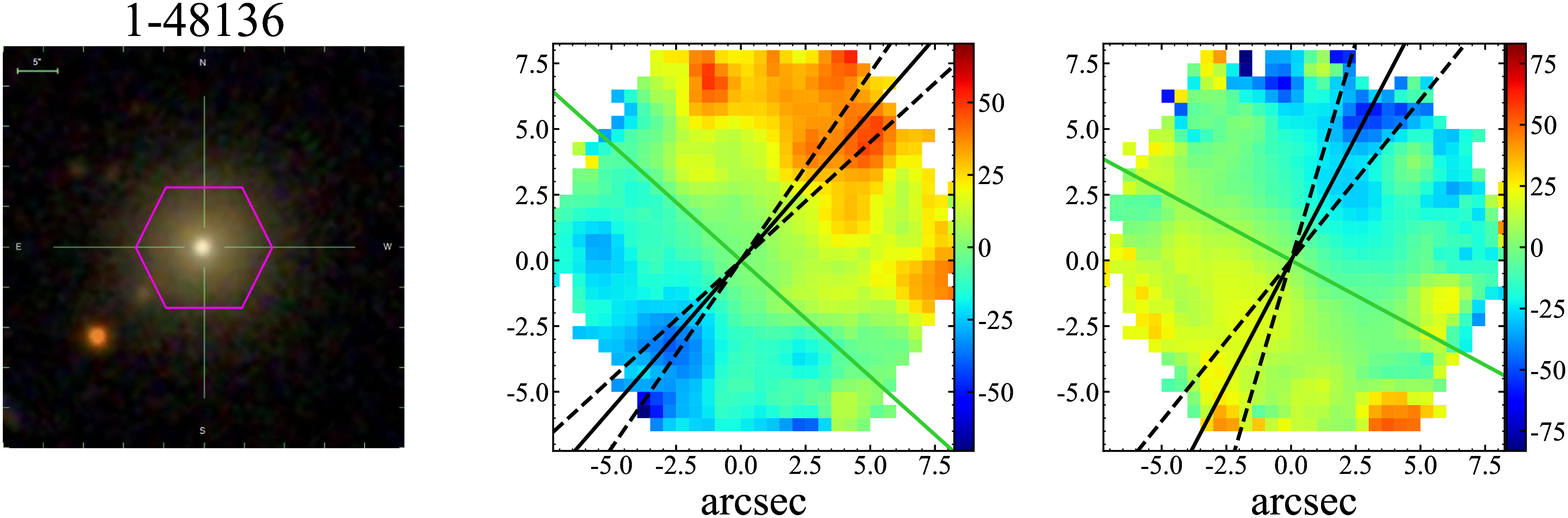}
    \caption{Examples of three MaNGA galaxies. Each row represents a galaxy (MaNGA ID is shown on the top). The left panel shows the SDSS \textit{g,r,i}-band image, the middle panel shows the stellar velocity field and the right panel shows the gas velocity field. The rotation velocities are indicated by the color bar, the red side is moving away from us while the blue side is approaching us. The solid black lines and green lines in each velocity field show the major and minor axis of the velocity field fitted by Python module FIT\_KINEMATIC\_PA \citep{krajnovic_kinemetry_2006}, while two dashed lines show $\pm 1 \sigma$ error range of PAs. The top row shows a galaxy (MaNGAID 1-55648) that has kinematically aligned stars and gas with $\Delta$PA $= 1^{\circ}$; the middle row shows a galaxy (MaNGAID 1-338633) with gas and stars rotating perpendicularly with $\Delta$PA $= 74^{\circ}$; the bottom row shows a galaxy (MaNGAID 1-48136) with gas and stars counter-rotating with $\Delta$PA $= 167^{\circ}$.}
    \label{fig:explain_PA}
\end{figure*}

In this paper, we select 460 gas-star misaligned galaxies from the internal Product Launch-10 (MPL-10) in Mapping Nearby Galaxies at Apache Point Observatory \citep[MaNGA,][]{2015ApJ...798....7B}, a new internal field spectroscopic survey. As a follow-up work of \cite{jin_sdss-iv_2016}, the sample size is enlarged by a factor of 7. Based on this largest misaligned sample so far, we look into the properties (i.e. kinematics, stellar populations, star formation activities, gas-phase metallicities) of these misaligned galaxies. The paper is organized as follows, In Section \ref{sec:data}, we present the selection of misaligned galaxies and their control samples. In Section \ref{sec:data analysis}, we compare the spatial resolved properties between misaligned galaxies and the control samples. We discuss the observational results in Section \ref{sec:discussion}. In Section \ref{sec:conclusions}, we briefly summarize the results. Throughout this paper, we adopt a flat $\Lambda$CDM cosmology with $\Omega_\Lambda=0.7$, $\Omega_{\rm m}=0.3$, and $H_0=70$ km s$^{-1}$ Mpc$^{-1}$.

\section{data}
\label{sec:data}

\subsection{The MaNGA survey}

MaNGA is one of the three core programs in the fourth-generation Sloan Digital Sky Survey (SDSS-IV) which began on July, 2014 \citep{2015ApJ...798....7B, drory_manga_2015}, it employs the Baryon Oscillation Spectroscopic Survey (BOSS) spectrographs \citep{smee_multi-object_2013} on the 2.5m Sloan Foundation Telescope \citep{gunn_25_2006}. The MaNGA observing strategy is described in \citet{law_observing_2015}, the MaNGA data proc pipeline is described in \citet{law_data_2016} and the flux calibration scheme is presented in \citet{yan_sdss-ivmanga_2016}. An overview of the survey execution strategy and data quality is provided in \citet{yan_sdss-iv_2016}. MaNGA has finished the survey of an unprecedented sample of $\sim$10,000 nearby galaxies early 2021 with a flat distribution in stellar masses in between $9 \le {\rm log}(M_*/M_\odot) \le 11$, the full redshift range is 0.01 $< z <$ 0.15 with a median value of $z \sim$ 0.03 \citep{2017AJ....154...86W, blanton_sloan_2017}. MaNGA employs dithered observations with 17 hexagonal integral field units (IFU) that vary in diameter from 12\arcsec (19 fibers) to 32\arcsec (127 fibers). Two dual-channel spectrographs provide simultaneous wavelength coverage over 3600$-$10,300 \r{A} with a median resolution $R \sim$ 2000. The typical spatial resolution is 1$\sim$2 kpc. The target galaxies are divided into "primary" and "secondary" sample following a ratio of 3 : 1. The primary sample is resolved out to $\sim$1.5 effective radius ($R_{\rm e}$, Petrosian 50\% light radius) while the secondary sample is observed to $\sim$2.5 $R_{\rm e}$. A total exposure time of 3 hours on-sky ensures a per-fiber \textit{r}-band continuum signal-to-noise ratio (S/N) per pixel of roughly 5 in the outskirts of target galaxies, with much higher S/N towards the center.

\subsection{Sample of kinematically misaligned galaxies}
\label{sec:Sample of kinematically misaligned galaxies}

\begin{figure}
    \includegraphics[width=\columnwidth]{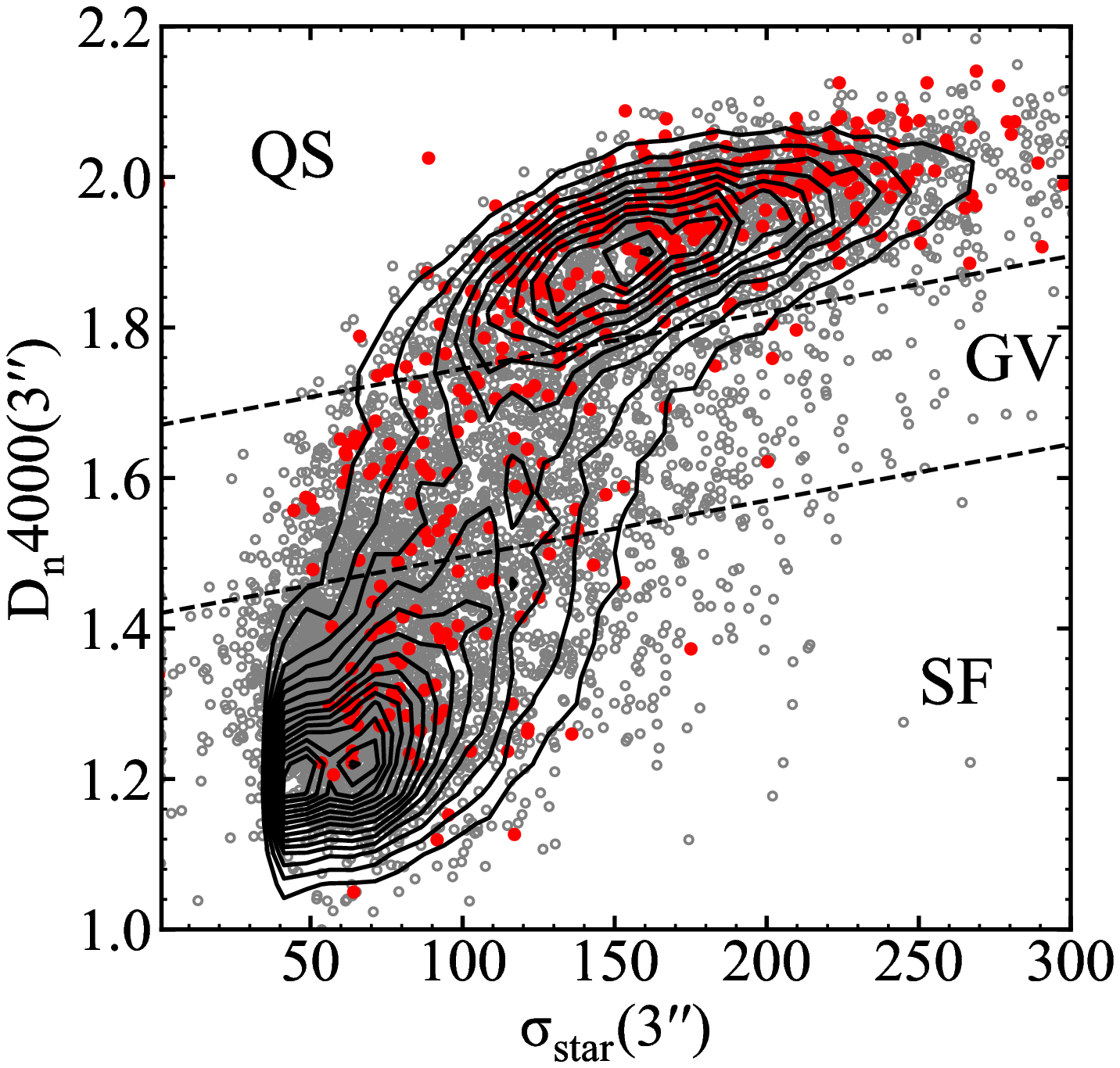}
    \caption{Diagram of central $3^{\prime \prime}$ D$_n$4000 versus ${\sigma}_{\rm star}$ for misaligned galaxies (red dots) and MaNGA sample (grey circles), with the SDSS DR7 sample shown as contours. There are two number density peaks in the contour. The two black dashed lines separate galaxies into star forming, green valley and quiescent sequence galaxies. The top dashed line is an approximation of the lower boundary of star forming main sequence (at the $\sim$1$\sigma$ level in scatter), while the bottom dashed line is an approximation of the upper boundary of quiescent sequence.}
    \label{fig:contour_Dn4000_V_DISP_3arcsec-3arcsec}
\end{figure}

\begin{table}
    \caption{Classification result of aligned/misaligned galaxies.}
    \label{tab:classification result}
    \begin{tabular}{lcc}
        \hline
        Aligned/Misaligned & Type & Number \\
        \hline
        Aligned & & 6502\\
        Misaligned & & \\
         & SF & 72\\
         & GV & 142\\
         & QS & 242\\
        Total & & 6958\\
        \hline
    \end{tabular}
\end{table}

\begin{figure}
    \includegraphics[width=1.0\columnwidth]{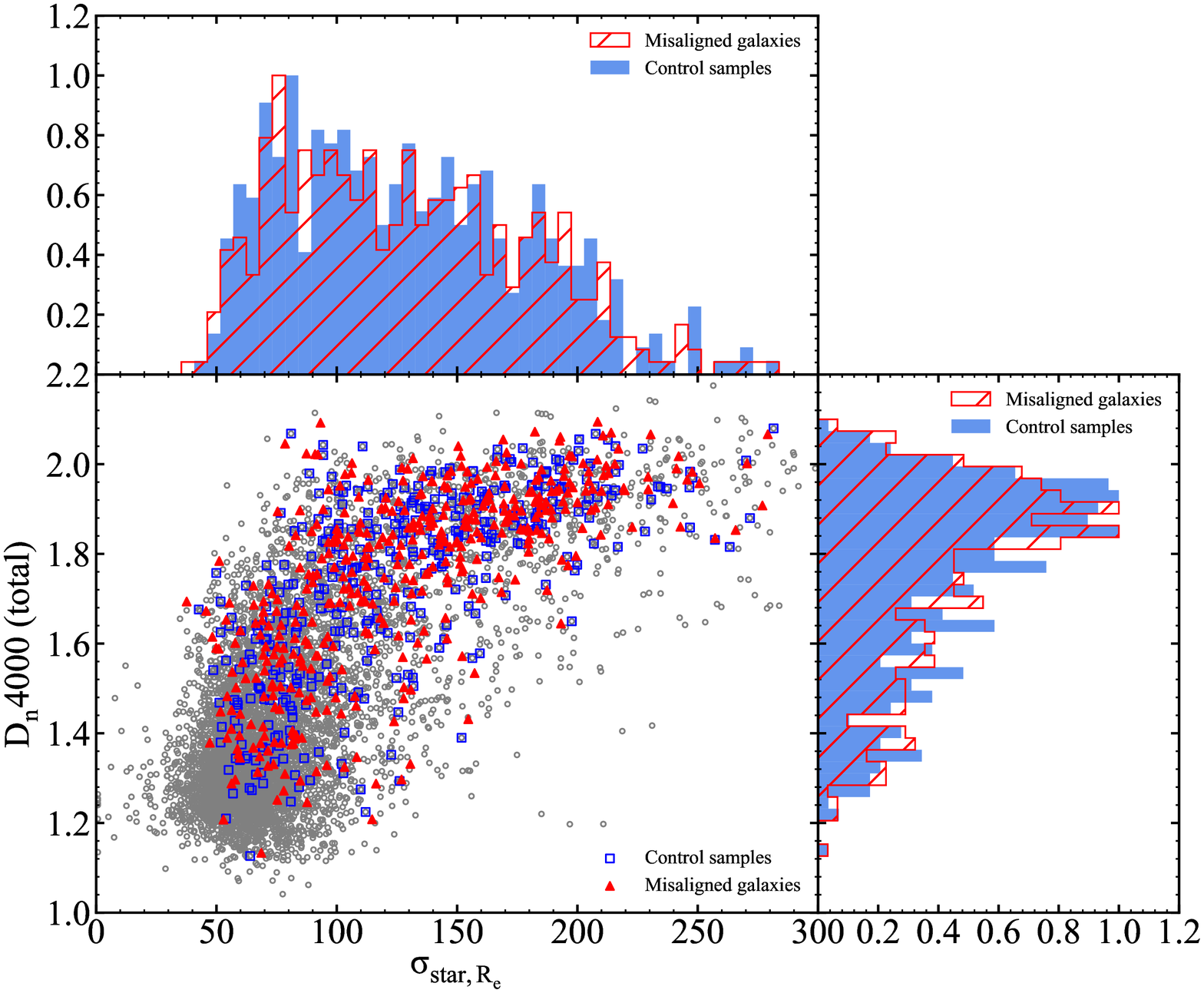}
    \caption{Total D$_n$4000 versus $\sigma_{{\rm star}, R_{\rm e}}$ for misaligned galaxies (red triangles) and one control sample (blue squares). The MaNGA galaxies are shown as grey circles. The top and right histograms show distribution of $\sigma_{{\rm star}, R_{\rm e}}$ and D$_n$4000, respectively. In each panel, the histograms filled with red lines are for the kinematically misaligned galaxies and that filled with blue color are for the control sample. The peaks of the distributions are set to 1.0.}
    \label{fig:Dn4000_sigma_star}
\end{figure}

The MaNGA sample and data products used in this work are drawn from the internal MaNGA Product Launch-10 (MPL-10), which includes 9456 unique galaxies. The MaNGA data analysis pipeline \citep[DAP,][]{westfall2019data} uses pPXF \citep{cappellari2004parametric} and a subset of stellar templates  from MaSTar library \citep{yan_sdss-iv_2019} to fit the stellar continuum in each spaxel. The data products include estimation of the stellar absorption and measurements of 21 major nebular emission lines in 3600$-$10,300 \r{A}, emission line fluxes are corrected for underlying stellar continuum absorption. The parameters we extracted from MaNGA DAP products named as "SPX-GAU-MILESCH-MASTARHC2", including : line-of-sight rotation velocity of stars ($V_{\rm star}$), stellar velocity dispersion (${\sigma}_{\rm star}$), line-of-sight rotation velocity of ionized gas ($V_{\rm gas}$), gas velocity dispersion (${\sigma}_{\rm gas}$), emission line fluxes (e.g. $\oiii\lambda$5007, $\ha$), lick indexes such as D$_n$4000, which is defined as the flux ratio between two narrow bands of 3850$-$3950 and 4000$-$4000 \r{A}.

In order to quantify the kinematic misalignment between stellar and gas components, we require the galaxies to have robust measurements of emission lines. We first separate the MaNGA sample into emission line galaxies and "line-less" galaxies."line-less" galaxies are defined as galaxies with $\ha$ signal-to-noise ratio smaller than 3 for more than $90\%$ spaxels within $\sim$1.5 $R_{\rm e}$. For 6958 emission line galaxies, we fit the kinematic position angle (PA) to both stellar (${\rm PA}_{\rm star}$) and gas (${\rm PA}_{\rm gas}$) velocity fields. The kinematic PA is measured based on established methods of \citet{krajnovic_kinemetry_2006}, it is defined as the counter-clockwise angle between north and a line that bisects the velocity field of gas or stars, measured on the receding side.

In Fig.~\ref{fig:explain_PA}, we show three examples of MaNGA galaxies. Each row represents a galaxy. The left panel shows the SDSS \textit{g,r,i}-band image (MaNGA ID is shown on the top), the middle panel shows the stellar velocity field and the right panel shows the velocity field of ionized gas traced by $\ha$. The values of rotation velocities are indicated by the color bar, the red side is moving away from us while the blue side is approaching us. The solid black and green lines in each velocity field show the kinematic major and minor axis fitted by Python module FIT\_KINEMATIC\_PA \citep{krajnovic_kinemetry_2006}, while two dashed lines show $\pm1\sigma$ error range. The top row shows a galaxy that has kinematically aligned gas and stellar components with $\Delta$PA = $1^{\circ}$; the middle row shows a galaxy with gas and stars rotating perpendicularly with $\Delta$PA $= 76^{\circ}$; the bottom row shows a counter-rotating galaxy with $\Delta$PA $= 167^{\circ}$. We identify 723 galaxies with robust PA measurement (error of PA < $60^{\circ}$) and $\Delta$PA $\ge 30^{\circ}$ as our kinematic decoupled sample. We eyeball both the photometry and velocity fields of these galaxies, removing irregular galaxies, broad line AGNs and on-going mergers. Finally the sample size is reduced to 456 galaxies.

Fig.~\ref{fig:contour_Dn4000_V_DISP_3arcsec-3arcsec} shows D$_n$4000-${\sigma}_{\rm star}$ relation for the central $3^{\prime \prime}$ of misaligned galaxies (red dots) and MaNGA galaxy sample (grey circles), with the SDSS DR7 sample shown as contours. The value of D$_n$4000 and ${\sigma}_{\rm star}$ within $3^{\prime \prime}$ are taken from MPA/JHU catalogue\footnote{https://wwwmpa.mpa-garching.mpg.de/SDSS/DR7/}. There are two density peaks in the contour. The peak at the bottom-left with lower D$_n$4000 and $\sigma_{\rm star}$ corresponds to younger stellar population, we refer them as star-forming (SF), the peak at the top-right with higher D$_n$4000 and $\sigma_{\rm star}$ corresponds to older stellar population, we refer them as quiescent sequence (QS). The population in between SF and QS is referred as green-valley (GV). The two black dashed lines separate these three populations. The top dashed line is an approximation of the lower boundary of star forming main sequence (at the $\sim$1$\sigma$ level in scatter), while the bottom dashed line is an approximation of the upper boundary of quiescent sequence. The classification result is listed in Table.~\ref{tab:classification result}. The reason that we do not apply the widely used SFR-$M_*$ relation to separate SF, GV, QS populations, includes: (1) different from SFR, D$_n$4000 can be measured consistently for all galaxies \citep{chen_post-starburst_2019}; (2) we hope to avoid any potential influence on the measurement of emission line strength due to the interaction between pre-existing and accreted external gas; (3) we find the median stellar velocity dispersion of the misaligned samples is 20$\sim$30 km$\rm \, s^{-1}$ larger than a control sample with similar D$_n$4000 and stellar mass distribution, considering that $\sigma_{\rm star}$ is a direct observed property that reflects the depth of the potential well of a galaxy \citep{2000ApJ...539L...9F, gebhardt2000relationship}, we use $\sigma_{\rm star}$ instead of $M_*$ in this work.

\subsection{Control sample}

It is believed that the gas components in the kinematically misaligned galaxies primarily originate from external processes like mergers and gas accretion. In order to understand the influence of external gas acquisition on the evolution of galaxies, we build control sample of galaxies with $\Delta$PA $< 30^{\circ}$. For each misaligned galaxy, we find ten non-misaligned galaxies which are closely matched in $\sigma_{{\rm star}, R_{\rm e}}$ ($|\Delta{\sigma}_{{\rm star}, R_{\rm e}}| \leq 8$) and global D$_n$4000 ($|\Delta$D$_n$4000$| \leq 0.05$) to construct the control sample. $\sigma_{{\rm star}, R_{\rm e}}$ is the stellar velocity dispersion at $R_{\rm e}$. Global D$_n$4000 is measured from the global spectrum within MaNGA bundle.

Fig.~\ref{fig:Dn4000_sigma_star} shows the misaligned sample (red triangles) and the control sample (blue squares) on the total D$_n$4000 versus $\sigma_{{\rm star}, R_{\rm e}}$ plane. The top and right histograms show distribution on $\sigma_{{\rm star}, R_{\rm e}}$ and D$_n$4000, respectively. It is clear that misaligned galaxies and control sample have the same distribution on $\sigma_{{\rm star}, R_{\rm e}}$ and total D$_n$4000.

Through quantifying the difference between the misaligned galaxies and the control sample, our aim is to have an idea about how the external processes influence galaxies evolution.

\section{data analysis}
\label{sec:data analysis}

\begin{figure*}
    \includegraphics[width=2.0\columnwidth]{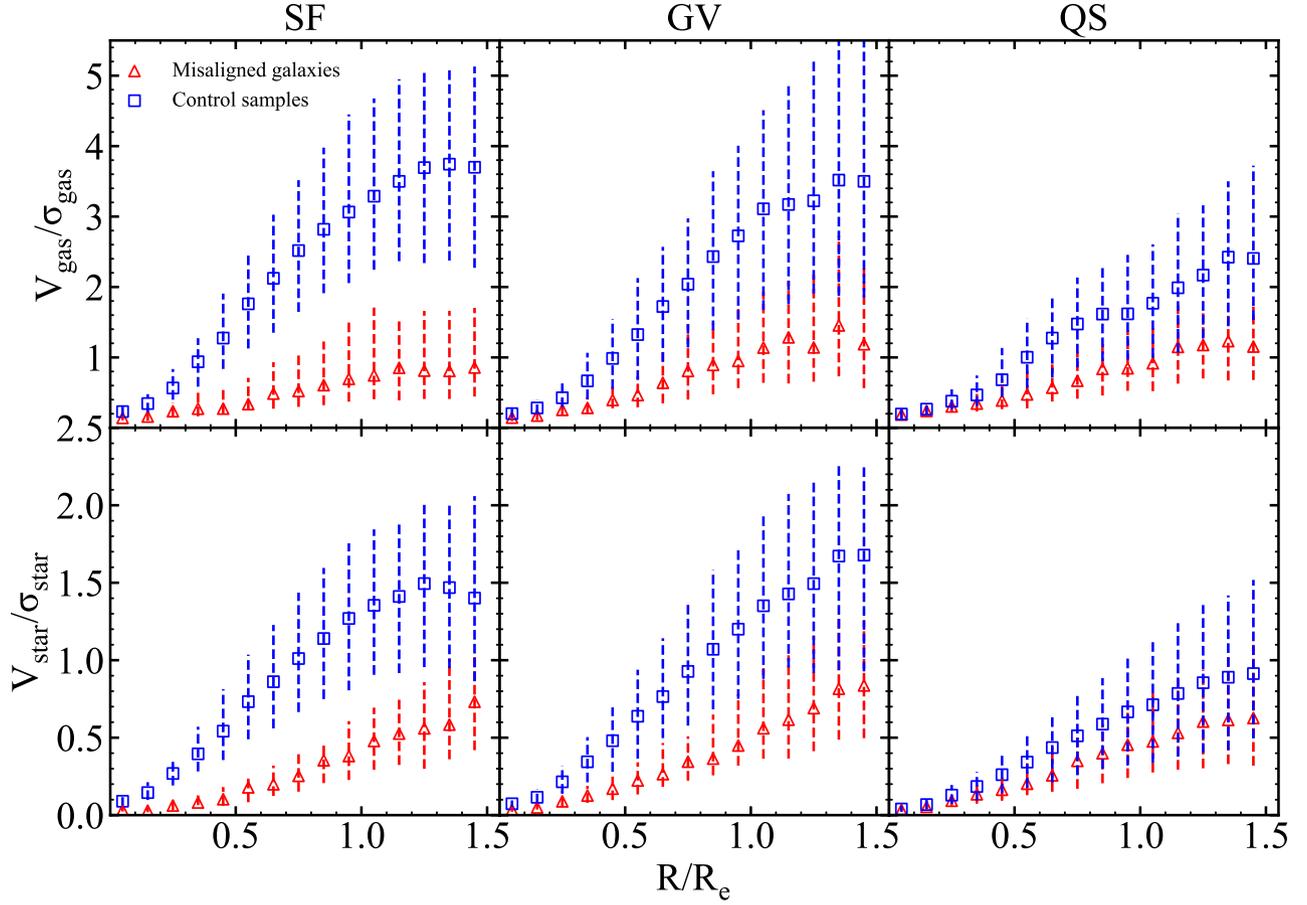}
    \caption{Comparison of ${\rm V}_{\rm gas}/{\sigma}_{\rm gas}$ (top row) and ${\rm V}_{\rm star}/{\sigma}_{\rm star}$ (bottom row) radial profiles within 1.5 ${\rm R}_{\rm e}$ between misaligned galaxies and control sample. The left panel is for SF, the middle panel is for GV, the right panel is for QS. The red triangles represent the median value of misaligned galaxies and the blue squares represent the median value of control sample. The error bars show the 30th and 70th percentiles of the distribution. The radii are the projected distances from the galactic centre to the spaxels where indices are measured in the unit of the effective radius.}
    \label{fig:vel_divide_sigma}
\end{figure*}

\begin{figure*}
    \includegraphics[width=2.0\columnwidth]{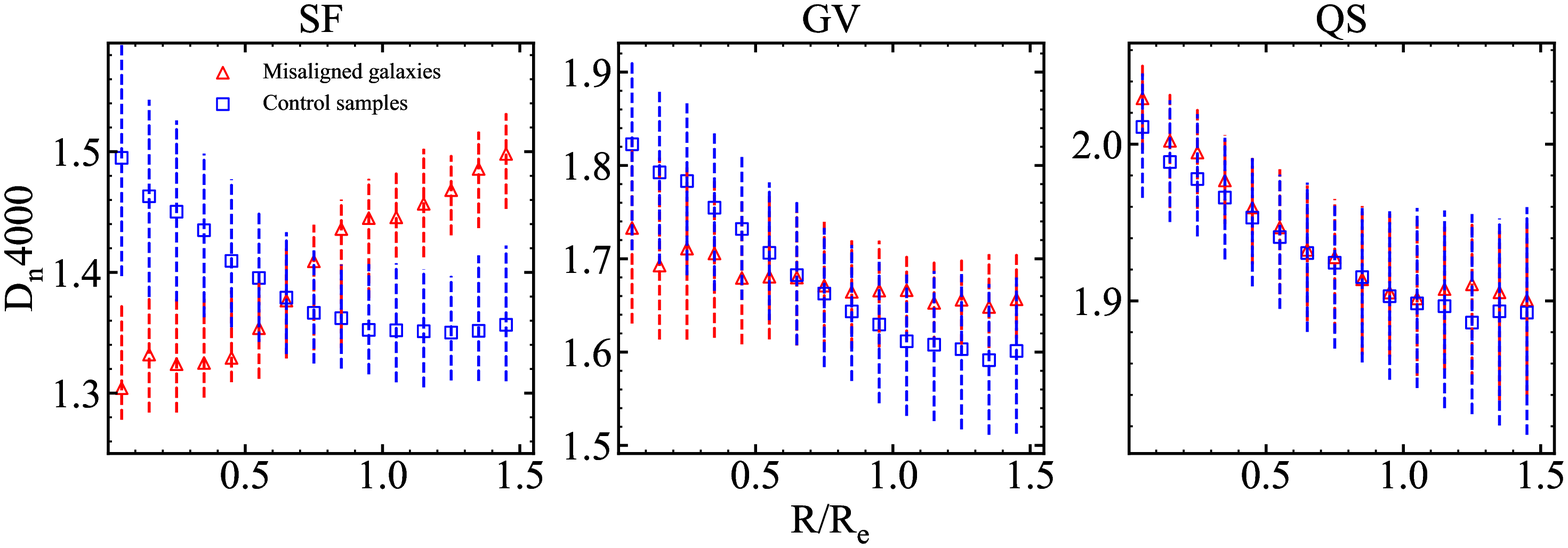}
    \caption{Comparison of D$_n$4000 radial profiles within 1.5 $R_{\rm e}$ between misaligned galaxies and control sample. Red triangles represent misaligned galaxies and blue squares represent control sample.}
    \label{fig:Dn(4000)}
\end{figure*}

In this section, we compare the spatial resolved properties between the misaligned galaxies and their control sample, including the average gas and stellar velocity to velocity dispersion ratio ($V_{\rm gas}/{\sigma}_{\rm gas}$ and $V_{\rm star}/{\sigma}_{\rm star}$), stellar population (D$_n$4000, light and mass-weighted stellar age, on going star forming rate), gas-phase metallicity. We hope to have a picture about the evolution of the misaligned galaxies through this section.

\subsection{Kinematics}

The ratio between ordered to random stellar motion in galaxies is strongly depending on luminosity and stellar mass \citep{veale_massive_2017, green_sami_2018}, which indicates a link between the build-up of stellar mass and angular momentum over cosmic time, which is fundamental in understanding the large variations in morphology and star formation in present-day galaxies. Major mergers are primary candidates for a dramatic changing in the morphology and spin of galaxies, however merger is only one of the many physical processes at play over the lifetime of a galaxy, continuing gas accretion and star formation can reshape the morphology and kinematics of remnants \citep{naab_atlas3d_2014}. In this section, we compare the ratio of radial gradients of velocity to velocity dispersion of gas and stellar components, $V_{\rm gas}/{\sigma}_{\rm gas}$ and $V_{\rm star}/{\sigma}_{\rm star}$, try to discern any difference in the formation/interaction history of the misaligned and control sample.

Fig.~\ref{fig:vel_divide_sigma} shows the median velocity to velocity dispersion ratio for gas and stellar components, $V_{\rm gas}/{\sigma}_{\rm gas}$ (top row) and $V_{\rm star}/{\sigma}_{\rm star}$ (bottom row), along their kinematic major axis for the misaligned sample and the control sample. In each row, the left panel is for SF, the middle panel is for GV and the right panel is for QS. The red triangles represent the median value of the misaligned sample and the blue squares represent the median value of the control sample. The error bars show the 30th and 70th percentiles of the distribution. The larger (smaller) value of $V/\sigma$ corresponds to more (less) rotationally support for the relevant components. $V$/$\sigma$ are taken from MaNGA DAP files, they are estimated from the spectral fitting process described in Sec.~\ref{sec:Sample of kinematically misaligned galaxies}. We can find that SF, GV, QS misaligned galaxies have $V_{\rm gas}/{\sigma}_{\rm gas}$ and $V_{\rm star}/{\sigma}_{\rm star}$ smaller and systematically shifted than their control samples. The difference between misaligned galaxies and control samples in QS is much smaller than that in SF and GV, indicating that external gas accretion has larger influence on the evolution (morphology) of SF, GV galaxies than QS galaxies.

\citet{chen_growth_2016} and \citet{jin_sdss-iv_2016} find that the central regions of star-forming misaligned galaxies show more intense, ongoing star formation and younger stellar populations than their outskirts, suggesting these galaxies accrete abundant external gas, the interaction between accreted and pre-existing gas triggers gas into central regions and forms new stars. The lower value of $V_{\rm star}/{\sigma}_{\rm star}$ and $V_{\rm gas}/{\sigma}_{\rm gas}$ in the star-forming misaligned galaxies than their controls can be easily understood under this picture in the following way: the interaction between accreted external gas and pre-existing gas (in the extreme case, they are counter-rotating) leads to the cancellation of angular momentum, thus lowers $V_{\rm gas}/{\sigma}_{\rm gas}$ as well as $V_{\rm star}/{\sigma}_{\rm star}$ once the gas transforms into stars. 

The GV and QS misaligned galaxies appear to have been undergoing a similar process to that seen in the star-forming misaligned galaxies which lead to the lower value of $V_{\rm star}/{\sigma}_{\rm star}$ and $V_{\rm gas}/{\sigma}_{\rm gas}$ than the control sample, but, on the one hand, the gas accretion and trigger of star formation happened earlier; on the other hand, the process is less violent in QS galaxies since the amount of both pre-existing and accreted gas is small in QS galaxies.

\begin{figure*}
    \includegraphics[width=2.0\columnwidth]{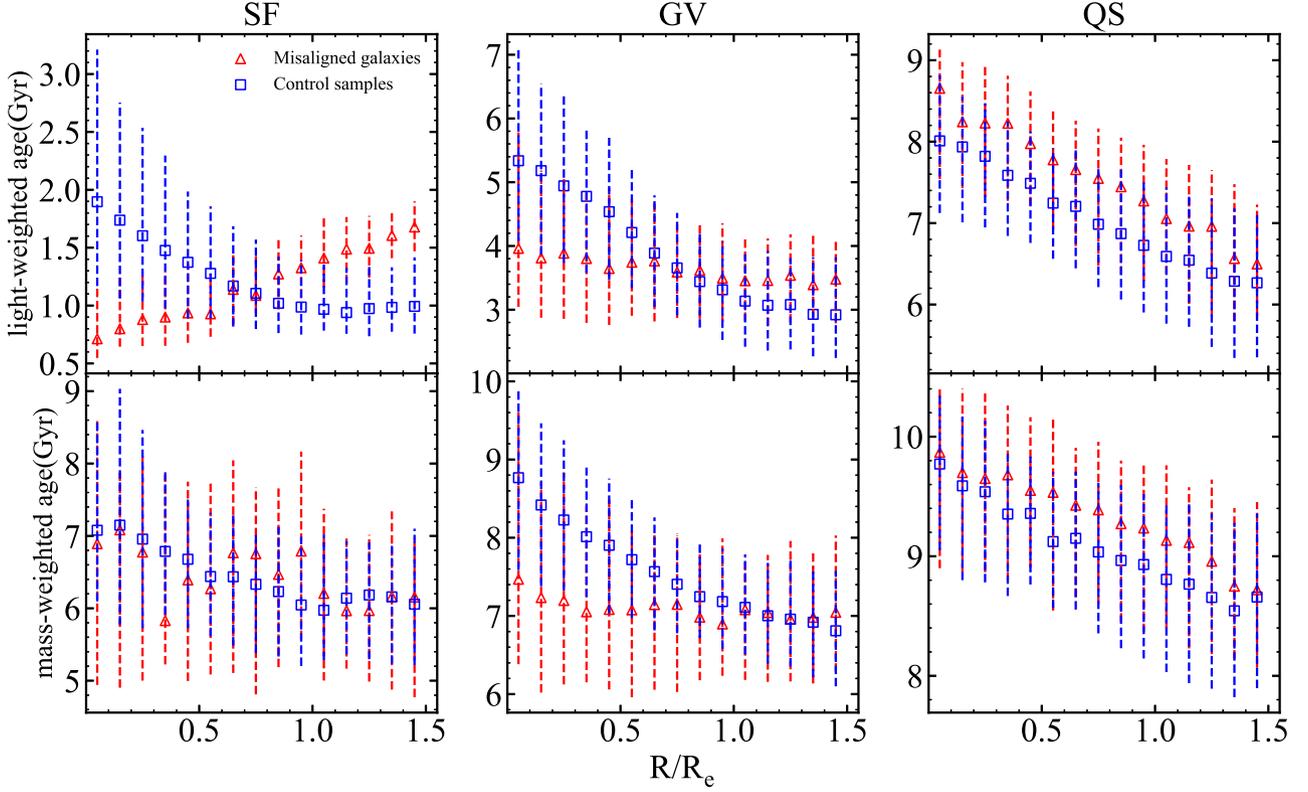}
    \caption{Comparison of light-weighted stellar ages and mass-weighted stellar age radial profiles within 1.5 $R_{\rm e}$ between misaligned galaxies and control sample. Red triangles represent misaligned galaxies and blue squares represent control sample.}
    \label{fig:lw_age_and_mw_age}
\end{figure*}

\begin{figure*}
    \includegraphics[width=1.6\columnwidth]{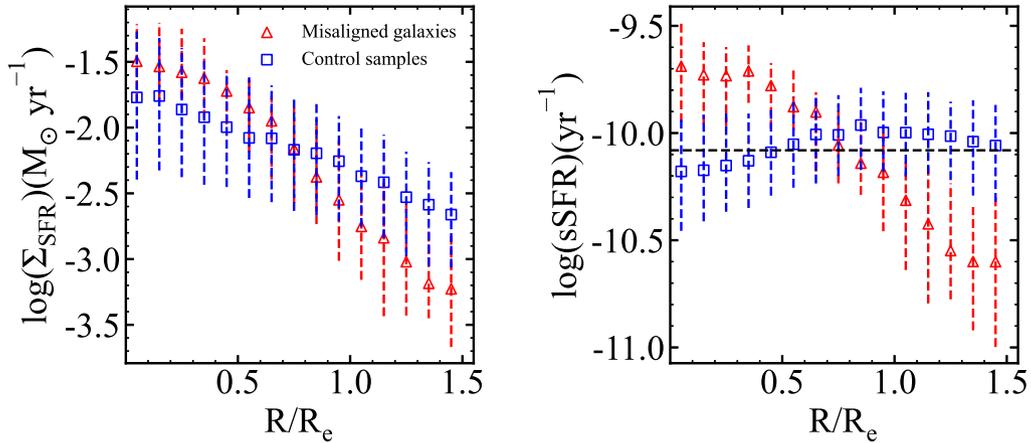}
    \caption{Comparison of star formation rate surface density and specific star formation rate radial profiles within 1.5 $R_{\rm e}$ between SF misaligned galaxies and control sample. Red triangles represent misaligned galaxies and blue squares represent control sample. The horizontal dashed line is defined as 1/($t_{\rm{H}(z)}-1\,\rm Gyr$), where $t_{\rm{H}(z)}$ is the Hubble time at the median redshift of the MaNGA star forming galaxy sample.}
    \label{fig:SFR_and_sSFR}
\end{figure*}

\subsection{Stellar population}

In this section, we study the stellar population distribution of the misaligned galaxies and control sample using continuum spectral indices D$_n$4000, light-weighted stellar age and mass-weighted stellar age as well as the ongoing star formation rate. 

Fig.~\ref{fig:Dn(4000)} shows the D$_n$4000 radial profiles for the SF (left), GV (middle) and QS (right) galaxies. Again, red triangles represent misaligned galaxies and blue squares represent control sample. For the star-forming galaxies, the D$_n$4000 gradient of misaligned galaxies is positive while the gradient of control sample is negative. At $R$ < $\sim$0.7 $R_{\rm e}$, the median value of D$_n$4000 is lower (younger in stellar population) for the misaligned galaxies than the control sample. Over this radius, the result is totally inverse, the stellar population in misaligned galaxies becomes older (higher D$_n$4000) than the control galaxies. This is consistent with the results in \citet{chen_growth_2016}, \citet{jin_sdss-iv_2016} and \citet{bizyaev_sdss_2019}, indicating that SF misaligned galaxies have younger stellar population in the centre than that in the outskirts. The explanation for this outside-in growth mode in SF misaligned galaxies is that the redistribution of angular momentum occurs from gas-gas collisions between the pre-existing and the accreted gas largely accelerates gas inflow, on the one hand leading to a fast centrally concentrated star formation, on the other hand shutting down the star formation in the outskirts due to the lack of cold gas. The controls have a negative gradient in D$_n$4000, indicating older stellar populations in the centre, as expected for ordinary bulge+disk structure of star-forming galaxies, consistent with the inside-out growth mode. For the GV misaligned galaxies (middle panel), the median D$_n$4000 has a flat distribution over radius, with a roughly constant value of 1.7, while control sample has a negative gradient in D$_n$4000. Similar to the SF galaxies, the stellar population is younger with lower D$_n$4000 in the misaligned galaxies than the control sample at $R$ < $\sim$0.7 $R_{\rm e}$, and older over this radius. This suggests that the GV galaxies was undergoing similar process after external gas accretion as the SF galaxies. The smaller difference in D$_n$4000 between GV misaligned galaxies and the control sample indicates that this process happened either earlier or less violent than SF galaxies. The misaligned QS and their control galaxies have identical negative D$_n$4000 gradients, indicating external gas accretion either happened much earlier than the SF and GV galaxies or the influence on stellar population is negligible. The difference in D$_n$4000 between the misaligned galaxies and control sample decreases from SF to QS.

Fig.~\ref{fig:lw_age_and_mw_age} shows radial profiles in light-weighted stellar age (top row) and mass-weighted stellar age (bottom row). The two ages are taken from the Pipe3D Value Added Catalog \citep{2016RMxAA..52...21S, 2016RMxAA..52..171S}, in which the stellar continuum is fitted as a linear combination of synthetic single stellar population (SSP) templates. Again red triangles represent misaligned galaxies and blue squares represent the control sample. Since D$_n$4000 is a good indicator of light-weighted age of a stellar population, the radial profiles of light-weighted age and D$_n$4000 are very similar. It enhances the difference between misaligned galaxies and control sample, which can be seen from the distinct difference of light-weighted age in QS while D$_n$4000 shows little difference in QS. Mass-weighted ages of the SF misaligned galaxies show an overall flat distribution of old stellar population along radius (6$\sim$7 Gyr), which is totally different from the positive gradient of the light-weighted ages. Similar to the SF misaligned galaxies, the mass-weighted age gradient of their control sample is very shallow, and has similar value to that of misaligned galaxies. This suggests the underlying stellar population of the star forming misaligned galaxies is similar to the control samples, the current central concentrated star formation is simply a recent burst on top of a dominant old population. The misaligned galaxies in GV have a median mass-weighted age of 7$\sim$8 Gyr across the whole galaxies. At $R$ < 0.7$\sim$0.8 $R_{\rm e}$, it is $\sim$1 Gyr younger than the control samples (middle panel of Fig.~\ref{fig:lw_age_and_mw_age}) and becomes consistent with the controls at outskirts. For the misaligned QS galaxies, we find that their mass weighted ages are consistent with the control samples over radius, there is weak evidence that the misaligned galaxies are a little bit older, we do not tend to explain this tiny difference considering that the error bar of mass-weighted age measurement is quite large.

We estimate star formation rate surface density ($\Sigma_{\rm SFR}$) of star forming regions from the dust extinction corrected $\ha$ luminosity \citep{salpeter_luminosity_1955, kennicutt_star_1998} : $\rm SFR(M_{\sun} \ yr^{-1}) = 7.9 \times 10^{-42} \ L_{\ha} (erg \ s^{-1})$. Dust extinction is calculated from Balmer decrement ($\ha$ and $\hb$ flux ratio) under Case B, the Galactic dust extinction curve from \citet{calzetti_dust_2001} is applied. 
The specific star formation rate (sSFR) is defined as SFR/$M_*$ for each spaxel.

Fig.~\ref{fig:SFR_and_sSFR} shows radial profiles of the star formation rate surface density (left panel) and specific star formation rate (right panel) for SF galaxies. Again red triangles represent misaligned galaxies and blue squares represent the control samples. We only calculated $\Sigma_{\rm SFR}$ and sSFR for SF galaxies since there are very few star forming spaxels in GV and QS galaxies. As we can see, misaligned galaxies have higher $\Sigma_{\rm SFR}$ and sSFR within 0.7$\sim$0.8 $R_{\rm e}$ than their control sample, indicating that misaligned galaxies have enhanced recent star forming activity in the central regions. The horizontal dashed line in the right panel of Fig.~\ref{fig:SFR_and_sSFR} is defined as 1/($t_{\rm{H}(z)}-1\,\rm Gyr$), where $t_{\rm{H}(z)}$ is the Hubble time at the median redshift of the MaNGA star forming galaxy sample, and 1 Gyr is subtracted to account for the fact star formation primarily occurred after reionization. The higher sSFR for the central region of the misaligned sample than the value marked by the horizontal dashed line indicates that current SFR of the misaligned galaxies are higher than the past average ($M_*/(t_{\rm{H}(z)}-1\,\rm Gyr)$) of the local star forming galaxies, while the control sample have a current SFR which is similar to the past average, indicating fast growth of the central regions in the misaligned galaxies triggered by external gas accretion.

\subsection{Gas-phase metallicity}

\begin{figure*}
    \includegraphics[width=2.0\columnwidth]{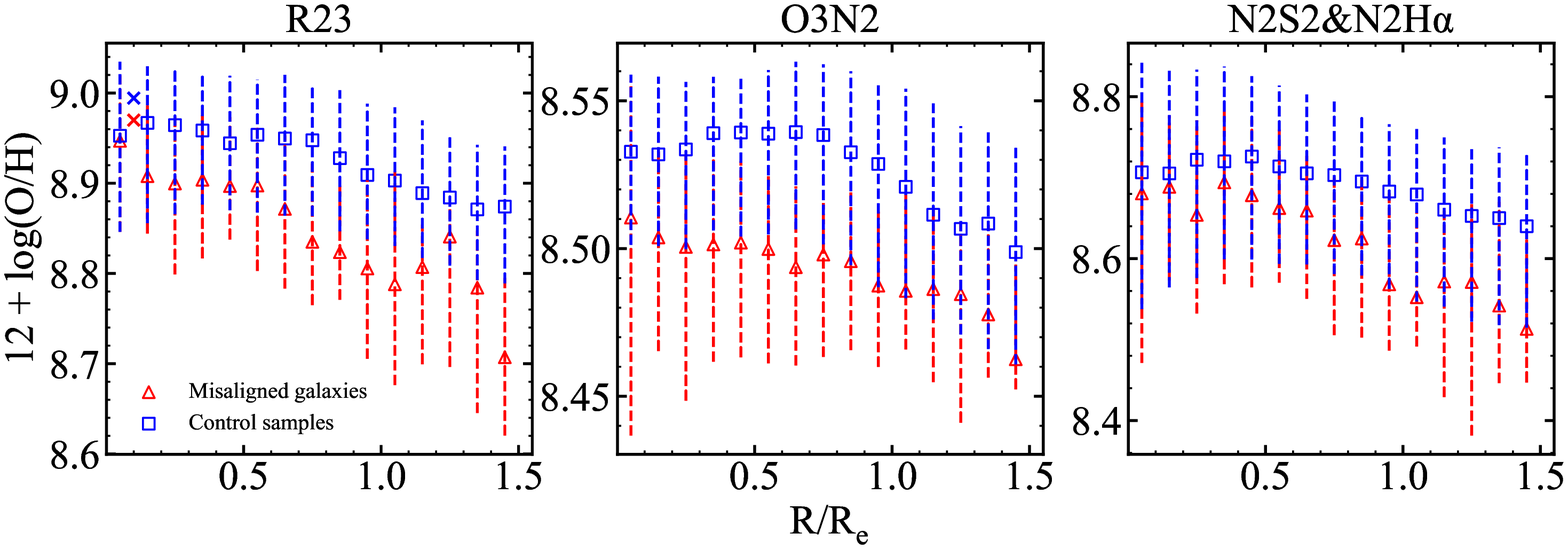}
    \caption{Comparison of the radial profiles of gas-phase metallicities estimated from three different strong-line calibrators. Red triangles represent SF misaligned galaxies and blue squares represent the control samples. The left panel of Fig.~\ref{fig:metallicity} applies $R_{23}$ as the metallicity calibrator (Eq.1 in \citet{tremonti_origin_2004}), the middle panel applies O3N2 as the metallicity calibrator (Eq.2 in \citet{marino2013o3n2}), while N2S2\&N2$\ha$ is used in the right panel (Eq.2 in \citet{dopita_chemical_2016}). Red and blue crosses in left panel mark the median value of central $3^{\prime \prime}$ gas-phase metallicity from MPA-JHU catalogue for misaligned galaxies and control samples.}
    \label{fig:metallicity}
\end{figure*}

\begin{figure*}
    \includegraphics[width=2.0\columnwidth]{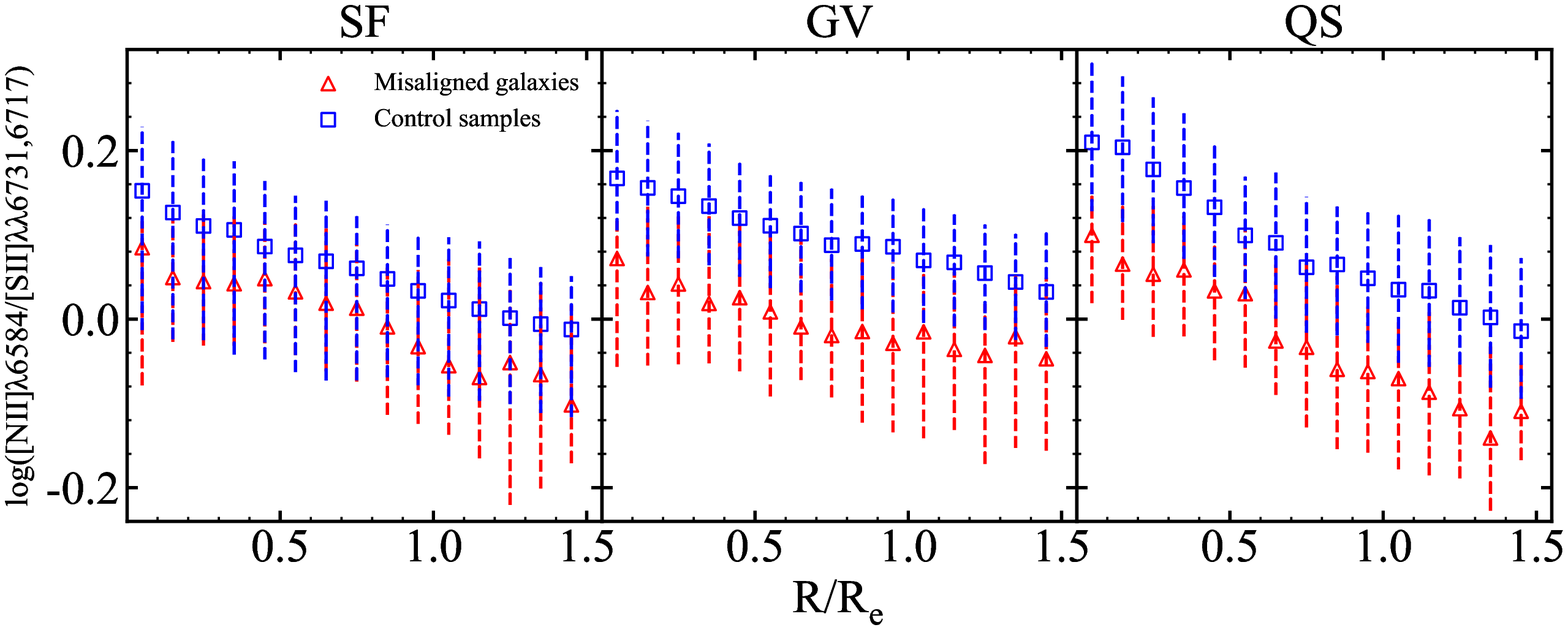}
    \caption{Comparison of {\nii}$\lambda$6584/{\sii}$\lambda\lambda$6717,6731 radial profiles within 1.5 $R_{\rm e}$ between misaligned galaxies and control sample. Red triangles represent misaligned galaxies and blue squares represent control sample.}
    \label{fig:NII_SII}
\end{figure*}

Gas-phase metallicity is one of the most fundamental physical properties of galaxies, it reflects the amount of gas reprocessed by stars and any exchange of gas between a galaxy and its environment.

The misaligned galaxies are believed to be the primary demonstrations whose evolution is regulated by external processes. External processes, such as minor/major mergers or gas accretion, could bring misaligned gas which has different metallicities from pre-existing gas into the galaxies. Thus, comparing the gas-phase metallicity between misaligned galaxies with their non-misaligned control sample will give us clues about the origin of misaligned gas as well as how the gas accretion processes influence the galaxy evolution.

Fig.~\ref{fig:metallicity} shows the radial profiles of gas-phase metallicity (estimated from three different strong line metallicity calibrators) for SF misaligned galaxies (red triangles) as well as their control samples (blue squares). The left panel of Fig.~\ref{fig:metallicity} applies $R_{23}$ as the metallicity calibrator (Eq.1 in \citet{tremonti_origin_2004}), the middle panel applies O3N2 as the metallicity calibrator (Eq.2 in \citet{marino2013o3n2}), while N2S2 and N2$\ha$ is used in the right panel (Eq.2 in \citet{dopita_chemical_2016}). Comparing the metallicities estimated from these three different calibrators, we find that although the absolute values and radial gradients of metallicity are different when we use different calibrators, they all point to the same conclusion that SF misaligned galaxies have overall lower gas-phase metallicity than their control samples. We do not discuss the difference in metallicities given by different calibrators in this work since unexpected systematic effect has been found by using different metallicity calibrators due to various reasons \citep{kewley_metallicity_2008, schaefer_sdss-iv_2019}.

Considering that the strong-line abundance diagnostics are developed based on the stellar population synthesis and photoionization models, it is limited to be only applied to $\hii$ regions \citep{kewley_using_2002}. However, the gas in the GV and QS galaxies is not excited by star formation for most of the spaxels, we thus follow \citet{jin_sdss-iv_2016} to apply $\nii\lambda6584/\sii\lambda\lambda6731,6717$ as an alternative gas-phase metallicity indicator. It is suggested by \citet{dopita_chemical_2016} and \citet{kashino2016hide} that $\nii\lambda6584/\sii\lambda\lambda6731,6717$ is a proxy for N/O ratio, and correlates with O/H abundance very well at 12 + log(O/H) $>$ 8.0. Also, the two emission lines are close in wavelength so the dust extinction correction is negligible. Fig.~\ref{fig:NII_SII} shows the radial gradients of $\nii\lambda6584/\sii\lambda\lambda6731,6717$ for the SF (left panel), GV (middle panel) and QS (right panel) misaligned galaxies (red triangles) as well as their control galaxies (blue squares). Similar to Fig.~\ref{fig:metallicity}, we find that the misaligned galaxies have lower gas-phase metallicities than their control samples, not only for the star forming galaxies, but also for green valley and quiescent galaxies, supporting the conclusion that the misaligned galaxies accreted low abundance gas from a gas-rich dwarf or cosmic web.

We have to point out that for the star forming misaligned galaxies, the difference in gas-phase metallicity between the misaligned galaxies and their controls strongly depends on how we define the control samples. We will discuss this in Section \ref{sec:discussion-1}.

\begin{figure*}
    \includegraphics[width=2.0\columnwidth]{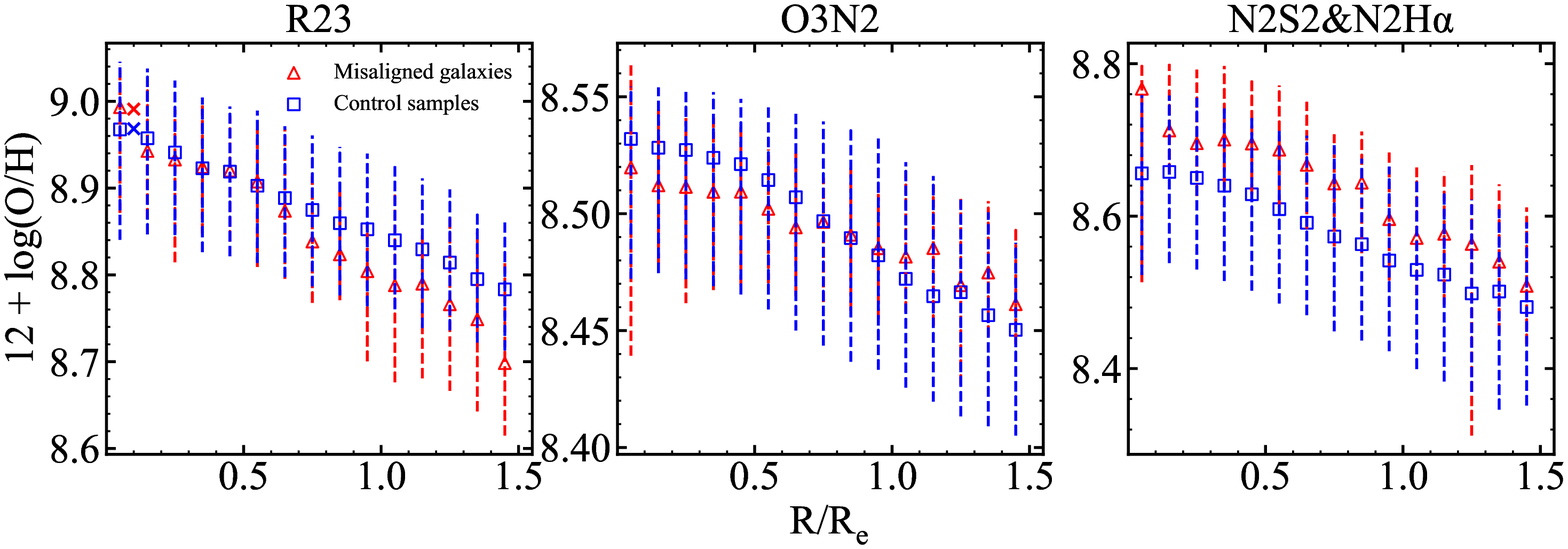}
    \caption{Comparison of gas-phase metallicities estimated from three different strong-line calibrators radial profiles within 1.5 $R_{\rm e}$ between misaligned galaxies and control sample closely matched in SFR and $M_*$. Same as Fig.~\ref{fig:metallicity}, but with the new control sample.}
    \label{fig:metallicity_SFR_M}
\end{figure*}

\begin{figure*}
    \includegraphics[width=2.0\columnwidth]{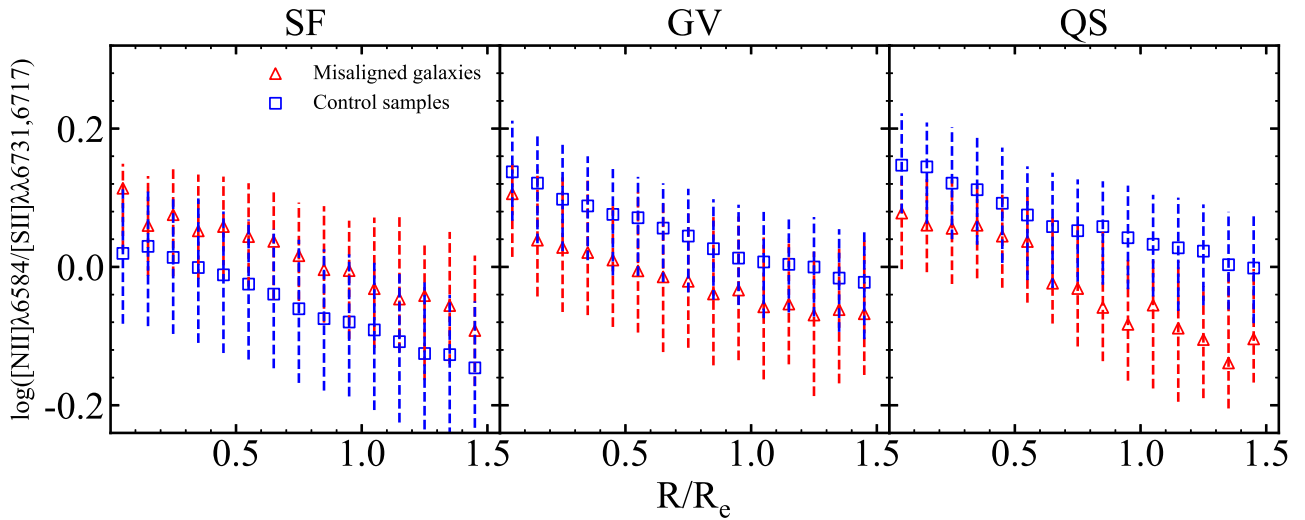}
    \caption{Comparison of {\nii}$\lambda$6584/{\sii}$\lambda\lambda$6717,6731 radial profiles within 1.5 $R_{\rm e}$ between misaligned galaxies and control sample closely matched in SFR and $M_*$. Same as Fig.~\ref{fig:NII_SII}, but with the new control sample.}
    \label{fig:NII_SII_SFR_M}
\end{figure*}

\begin{figure}
    \includegraphics[width=1.0\columnwidth]{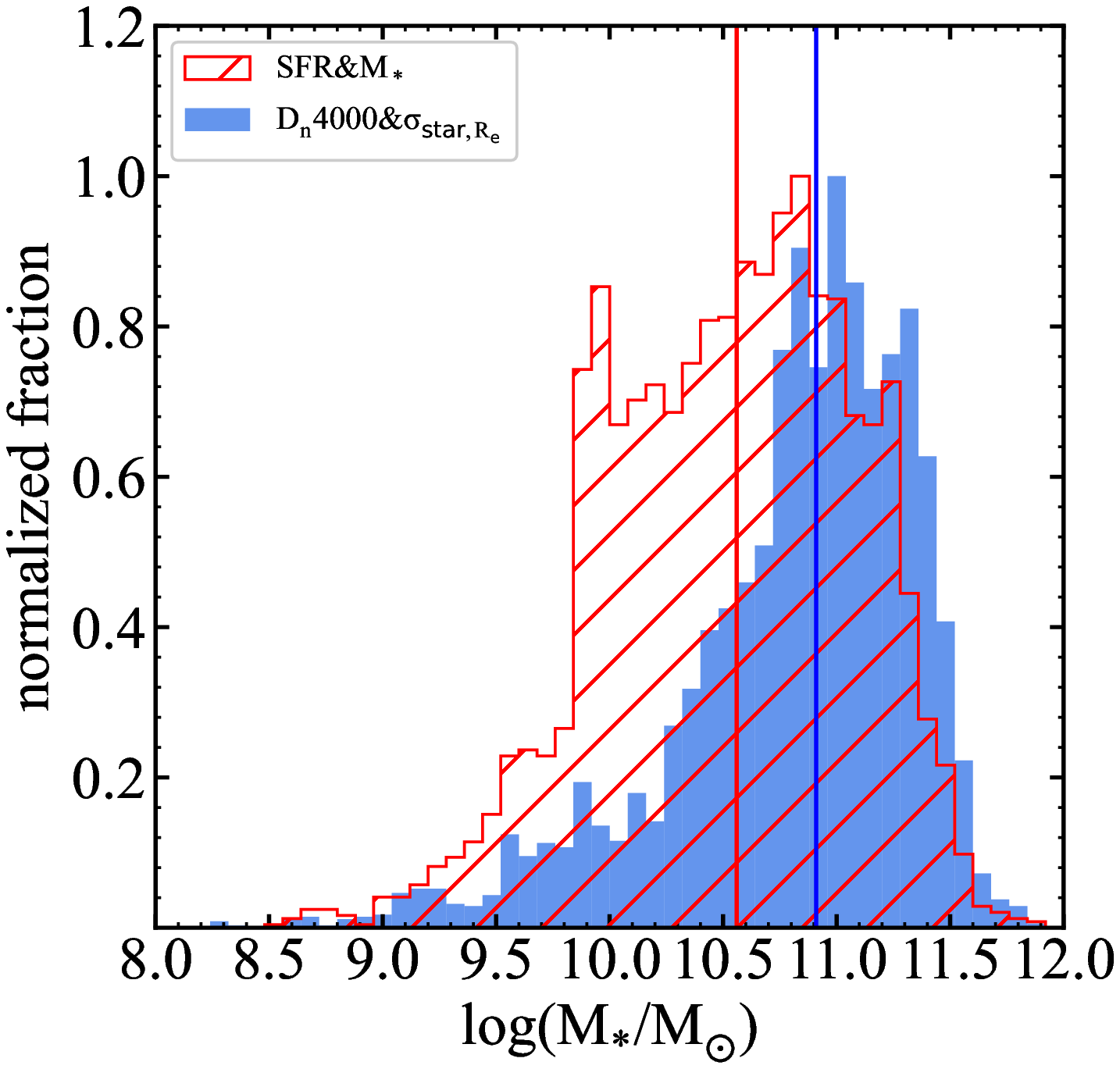}
    \caption{The $M_*$ distribution of control sample closely matched in SFR, $M_*$ (red histograms) as well as control sample closely matched in D$_n$4000, $\sigma_{{\rm star}, R_{\rm e}}$ (blue histograms). The red line is the median $M_*$ of control sample closely matched in SFR, $M_*$ while the blue line is the median $M_*$ of control sample closely matched in D$_n$4000, $\sigma_{{\rm star}, R_{\rm e}}$. The peaks of histogram are set to 1.0.}
    \label{fig:diff_M}
\end{figure}

\section{Discussion}
\label{sec:discussion}

Many previous works have studied the gas-star misalignment phenomena mostly in early type galaxies but rarely in star forming galaxies. Thanks to the large MaNGA sample, \citet{chen_growth_2016} and \citet{jin_sdss-iv_2016} detected gas-star misalignment in star forming galaxies. As a following work of \citet{jin_sdss-iv_2016}, we study the kinematics, stellar populations, star formation activities as well as metallicity of misaligned galaxies, and try to figure out the formation scenarios of misaligned galaxies, as well as how these mechanisms affect the properties of galaxies.

\subsection{Discrepancy in metallicity of SF misaligned galaxies}
\label{sec:discussion-1}
Our results on gas-phase metallicity of GV and QS misaligned galaxies are consistent with previous works \citep{chen_growth_2016, jin_sdss-iv_2016} that GV and QS misaligned galaxies have lower gas-phase metallicity than control samples. At first sight the result on gas-phase metallicity in SF misaligned galaxies seems contrary to previous works. \citet{chen_growth_2016} and \citet{jin_sdss-iv_2016} found SF misaligned galaxies have higher gas-phase metallicity than typical stellar mass-metallicity relation for local star-forming galaxies \citep{tremonti_origin_2004}, but this work finds SF misaligned galaxies have lower gas-phase metallicity than control samples. In this section, we focus on understanding the different results in gas-phase metallicity of SF misaligned galaxies and the dependance of our results on the selection of control samples.

We select another set of ten control samples that are closely match in SFR and $M_*$ with $|\Delta{\rm \log{SFR}}| \leq 0.2$ and $|\Delta{\log{M_*}}| \leq 0.1$, we repeat Fig.~\ref{fig:vel_divide_sigma} to Fig.~\ref{fig:NII_SII} based on the new control samples, finding that all the results are consistent with previous sections except the gas-phase metallicity of SF misaligned galaxies. Fig.~\ref{fig:metallicity_SFR_M} shows gas-phase metallicity estimated from three different strong-line calibrators for misaligned galaxies and the new control samples. The metallicity of the central region traced by $R_{23}$, N2S2\&N2$\ha$ and $\nii\lambda6584/\sii\lambda\lambda6731,6717$ is higher in misaligned galaxies than control samples, which is opposite to Fig.~\ref{fig:metallicity} and \ref{fig:NII_SII}. The central metallicity estimated from O3N2 of misaligned galaxies is still slightly lower than that of control sample. The red and blue crosses in the left panel of Fig.~\ref{fig:metallicity_SFR_M} mark the median value of central $3^{\prime \prime}$ gas-phase metallicity from MPA-JHU catalogue for misaligned galaxies and control samples. Our gas-phase metallicity in the central region is totally consistent with that from MPA-JHU catalogue. The central metallicity estimated from $R_{23}$ is $\sim$0.03 dex higher in the SF misaligned galaxies than control samples selected by SFR, $M_*$, but $\sim$0.03 dex lower than control samples closely matched in total D$_n$4000, $\sigma_{{\rm star}, R_{\rm e}}$ (left panel of Fig.~\ref{fig:metallicity}).

In order to understand the different results of gas-phase metallicity, we investigate the properties of different control samples. Fig.~\ref{fig:diff_M} shows the $M_*$ distribution of the two sets of control samples, it is clear that the $M_*$ of control samples selected by total D$_n$4000 and $\sigma_{{\rm star}, R_{\rm e}}$ is $\sim$0.35 dex larger than that of control sample selected by SFR, $M_*$. This $\sim$0.35 dex difference in $M_*$ is large enough to explain the $\sim$0.06 dex difference in metallicity between the two sets of control samples. SF misaligned galaxies do not follow the typical $M_*-\sigma_{{\rm star}, R_{\rm e}}$ relation. They have smaller $M_*$ than control sample closely matched in $\sigma_{{\rm star}, R_{\rm e}}$, but have larger $\sigma_{{\rm star}, R_{\rm e}}$ than control sample closely matched in $M_*$. Why the SF misaligned galaxies are the outliers in the $M_*-\sigma_{{\rm star}, R_{\rm e}}$ relation, one possibility is that the newly formed stars inherited the angular momentum of accreted gas that was not consistent with pre-existing stars, the contribution of these two parts of stars could broaden the observed stellar velocity dispersion, similar to 2$\sigma$ galaxies \citep{krajnovic_atlas3d_2011}. Limited by MaNGA spectral resolution, we can not clearly distinguish multiple stellar components with different line of sight velocity, future observations of higher spectral resolution are required to figure out this possibility.

\subsection{Formation scenarios of misaligned galaxies}
\label{sec:discussion-2}

In the last decades, several formation scenarios of gas-star misalignments have been proposed, including external gas accretion from dwarf companions as well as cosmic webs \citep{1998ApJ...506...93T, chen_growth_2016, jin_sdss-iv_2016}, minor/major mergers \citep{naab_atlas3d_2014, li_impact_2020}. Recent simulations \citep{starkenburg_origin_2019, duckworth_decoupling_2020, koudmani_little_2021} find a higher incidence of AGN in misaligned galaxies, suggesting AGN feedback also plays a role in the formation of misalignment.

In this work, we suggest that from SF to QS, these different types of galaxies have survived from different formation scenarios. \citet{chen_growth_2016} and \citet{jin_sdss-iv_2016} proposed that in the SF misaligned galaxies, the interaction between accreted gas and pre-existing gas leads to the consumption of angular momentum, large amount of gas flows to the central regions of galaxies, triggering the central star formation/starburst. Our results for SF galaxies in Fig.~\ref{fig:vel_divide_sigma} to Fig.~\ref{fig:SFR_and_sSFR} totally support this picture. The lower $V_{\rm gas}/{\sigma}_{\rm gas}$ and $V_{\rm star}/{\sigma}_{\rm star}$ of the misaligned galaxies than the control samples are due to angular momentum loss, lower D$_n$4000 and light-weighted ages, as well as higher $\Sigma_{\rm SFR}$ and sSFR at the central regions of SF misaligned galaxies are consistent with central star formation. We suggest that the primary formation mechanisms for SF misaligned galaxies is external gas accretion. The results for GV misaligned galaxies from Fig.~\ref{fig:vel_divide_sigma} to Fig.~\ref{fig:lw_age_and_mw_age} show similar trend as SF misaligned galaxies, suggesting the GV misaligned galaxies have undergone similar formation mechanisms as the star forming misaligned galaxies. On the other hand, most SF and GV misaligned galaxies show spiral or S0-like morphology. Numerical simulations suggest that mergers, especially major mergers can hardly keep the disk structure in spiral or S0 galaxies \citep{1992ApJ...393..484B} except in some extreme conditions, such as a major merger with a spiral-in falling \citep{zeng_formation_2021}.

For the QS misaligned galaxies, they show obvious difference from SF and GV galaxies. Although QS misaligned galaxies also have lower $V_{\rm gas}/{\sigma}_{\rm gas}$ and $V_{\rm star}/{\sigma}_{\rm star}$ than the control galaxies, the difference is much smaller than that in SF, GV. There is evidence that the QS misaligned galaxies have older stellar populations than the control samples as suggested by D$_n$4000, light- and mass-weighted ages. We suggest there are three different formation mechanisms for QS misaligned galaxies: (1) QS misaligned galaxies are the evolutionary results of GV galaxies; (2) merger contribution is larger in QS misaligned galaxies than SF and GV misaligned galaxies. \citet{li_impact_2020} find merger fraction is $\sim$10\% higher in QS misaligned galaxies than the co-rotating counterparts, while the difference in merger fraction between SF, GV misaligned galaxies and their co-rotating counterparts is not obvious; (3) the QS misaligned galaxy is formed through gas accretion by a gas-poor elliptical galaxy and this gas accretion process does not influence the evolution of progenitors. In this case, progenitors with less gas, much older stellar populations are easier to show misaligned phenomena since the interaction between accreted and pre-existing gas can be neglected, it is easier for the accreted gas to keep their original angular momentum. In this scenario, the observed difference between QS misaligned galaxies and their control galaxies is totally due to the difference between the progenitors.

A new class of elliptical galaxy termed as `red geyser' has been identified from MaNGA \citep{2016Natur.533..504C}. These `red geysers' are believed to exhibit high velocity outflow triggered by AGN. And the misaligned phenomenon is prevalence in `red geysers'. Based on cosmological simulations, \citet{starkenburg_origin_2019}, \citet{duckworth_decoupling_2020} and \citet{ koudmani_little_2021} investigated the impact of AGN feedback on the formation of low-mass misaligned galaxies, suggesting that low-mass misaligned galaxies tend to have increased AGN feedback and increased energy injection, which leads to the removal of pre-existing gas and further reduces the interaction between accreted and pre-existing gas, follow up external gas accretion leads to the gas-star misaligned phenomenon. In this work, we also find that the AGN fraction is 2-3 times higher in the SF and GV misaligned galaxies than their control samples, however, we would like to trust this as a nature result of gas inflow. The gas inflow on the one hand triggered the central star formation, on the other hand provided fuels for the central BHs, leading to a higher AGN fraction. The gas removal scenario suggested by the literatures can not explain the enhanced central star formation in SF and GV misaligned galaxies. For the QS misaligned galaxies, we do not find any obvious difference in AGN fraction between the misaligned and control samples.

\section{Conclusions}
\label{sec:conclusions}

We identify galaxies with kinematically misaligned gas and stellar components from the internal MaNGA Product Launch-10, generating a sample of 72 star-forming galaxies, 142 green-valley galaxies, 242 quiescent galaxies. For each misaligned galaxies, we select ten control galaxies with similar $\sigma_{{\rm star}, R_{\rm e}}$ and total D$_n$4000 to build ten control samples. Through comparing the spatial resolved properties, including kinematics, stellar populations and gas-phase metallicities, between the misaligned galaxies and their control samples, we find that:

i) the misaligned galaxies have lower values in $V_{\rm gas}/{\sigma}_{\rm gas}$ and $V_{\rm star}/{\sigma}_{\rm star}$ (the ratio between ordered to random motion of gas and stellar components) than their control samples, and the difference between QS misaligned galaxies and their controls are much smaller than that in SF and GV galaxies.

ii) both SF and GV misaligned galaxies have younger stellar populations in their central regions with $R$ < $\sim$0.7 $R_{\rm e}$ than the control samples as indicated by D$_n$4000 as well as light-weighted ages. But over this radius, the control samples become younger than the misaligned galaxies, suggesting an outside-in growth mode in SF and GV misaligned galaxies. The difference in D$_n$4000 between the misaligned QS galaxies and their controls is tiny, but the light-weighted ages enhances the difference between misaligned galaxies and the control samples, showing that the misaligned galaxies are a little bit older. The SF misaligned galaxies show higher star formation rate surface density ($\Sigma_{\rm SFR}$) and specific star formation rate at $R$ < 0.7$\sim$0.8 $R_{\rm e}$ than the control samples, suggesting enhanced central concentrated star formation.

iii) different from the light-weighted ages, the mass-weighted ages are dominated by the older and lower mass stars. The SF misaligned galaxies show an overall flat distribution in mass-weighted ages of a value 6$\sim$7 Gyr, this radial distribution is similar to their control samples, suggesting that the enhanced current central concentrated star formation suggested by D$_n$4000, light-weighted ages, $\Sigma_{\rm SFR}$ and sSFR is simply a recent burst on top of a dominant old population. The misaligned GV galaxies have a median mass-weighted age of 7$\sim$8 Gyr over the entire galaxies, and it is $\sim$1 Gyr younger than the control samples at $R$ < 0.7$\sim$0.8 $R_{\rm e}$. The difference in mass-weighted ages between QS misaligned galaxies and their control is tiny.

iv) three different strong-line metallicity calibrators as well as $\nii\lambda6584/\sii\lambda\lambda6731,6717$ are used to traced the gas-phase metallicity of the misaligned galaxies and their control samples, all of them give a similar result that misaligned galaxies have lower gas-phase metallicity than control samples.

Combining all these observational results, we suggest that SF misaligned galaxies formed through accreting external gas from a gas-rich dwarf or cosmic web, redistribution of angular momentum happens through the interaction between the pre-existing and accreted gas, triggering gas inflows and follow up fast central concentrated star formation. The GV misaligned galaxies have survived similar process as that in SF misaligned galaxies. While the QS misaligned galaxies have survived from different formation scenarios.

\section*{Acknowledgements}

Y. C acknowledges support from the National Key R\&D Program of China (No. 2017YFA0402700), the National Natural Science Foundation of China (NSFC grants 11573013, 11733002, 11922302), the China Manned Space Project with NO. CMS-CSST-2021-A05. The authors are very grateful to the referee for valuable comments and suggestions which improved the work a lot.

Funding for the Sloan Digital Sky Survey IV has been provided by the Alfred P.
Sloan Foundation, the U.S. Department of Energy Office of Science, and the Participating Institutions.
SDSS-IV acknowledges support and resources from the Center for High-Performance Computing at
the University of Utah. The SDSS web site is www.sdss.org.

SDSS-IV is managed by the Astrophysical Research Consortium for the Participating Institutions of the SDSS Collaboration 
including the Brazilian Participation Group, the Carnegie Institution for Science, Carnegie Mellon University, the Chilean Participation 
Group, the French Participation Group, Harvard-Smithsonian Center for Astrophysics, Instituto de Astrof\'isica de Canarias, The Johns 
Hopkins University, Kavli Institute for the Physics and Mathematics of the Universe (IPMU) / University of Tokyo, Lawrence Berkeley 
National Laboratory, Leibniz Institut f\"ur Astrophysik Potsdam (AIP), Max-Planck-Institut f\"ur Astronomie (MPIA Heidelberg), 
Max-Planck-Institut f\"ur Astrophysik (MPA Garching), Max-Planck-Institut f\"ur Extraterrestrische Physik (MPE), National 
Astronomical Observatories of China, New Mexico State University, New York University, University of Notre Dame, 
Observat\'ario Nacional / MCTI, The Ohio State University, Pennsylvania State University, Shanghai Astronomical Observatory, 
United Kingdom Participation Group, Universidad Nacional Aut\'onoma de M\'exico, University of Arizona, University of Colorado Boulder, 
University of Oxford, University of Portsmouth, University of Utah, University of Virginia, University of Washington, University of Wisconsin, 
Vanderbilt University, and Yale University.

\section*{Data Availability}
The data underlying this article will be shared on reasonable request to the corresponding author.

\bsp	
\label{lastpage}
\end{document}